%% file: ms.tex
\documentclass[12pt,preprint]{aastex}   

\slugcomment{Astronomical Journal, in press}

\shorttitle{Hot subdwarfs rejected from the PG catalog: an assessment}
\shortauthors{Wade, Stark, Green \& Durrell}

\begin{document}

\title{HOT SUBDWARF STARS AMONG THE OBJECTS REJECTED 
FROM THE PG CATALOG: A FIRST ASSESSMENT USING GALEX PHOTOMETRY}

\author{Richard A.\ Wade and M.\ A.\ Stark}
\affil{Department of Astronomy and Astrophysics, The Pennsylvania
State University, 525 Davey Lab, University Park, PA 16802}
\email{wade@astro.psu.edu, stark@astro.psu.edu}

\author{Richard F.\ Green}
\affil{Large Binocular Telescope Observatory, University of Arizona, 
933 N. Cherry Ave., Tucson, AZ 85721-0065}
\email{rgreen@as.arizona.edu}

\author{Patrick R.\ Durrell}
\affil{Department of Physics and Astronomy, Youngstown State University, 
Youngstown, OH}
\email{prdurrell@ysu.edu}

\begin{abstract}
The hot subdwarf (sd) stars in the Palomar Green (PG) catalog of
ultraviolet excess (UVX) objects play a key role in investigations of
the frequency and types of binary companions and the distribution of
orbital periods. These are important for establishing whether and by
which channels the sd stars arise from interactions in close binary
systems.  It has been suggested that the list of PG sd stars is biased
by the exclusion of many stars in binaries, whose spectra show the
\ion{Ca}{2} K line in absorption.  A total of 1125 objects that were
photometrically selected as candidates were ultimately rejected from the
final PG catalog using this K--line criterion.  We study 88 of these
``PG--Rejects" (PGRs), to assess whether there are significant numbers of
unrecognized sd stars in binaries among the PGR objects.  The presence
of a sd should cause a large UVX, compared with the cool K--line star.
We assemble GALEX, Johnson $V$, and 2MASS photometry and compare the
colors of these PGR objects with those of known sd stars, cool single
stars, and hot+cool binaries.  Sixteen PGRs were detected in both the
far-- and near--ultraviolet GALEX passbands. Eleven of these, plus the 72
cases with only an upper limit in the far--ultraviolet band, are
interpreted as single cool stars, appropriately rejected by the PG
spectroscopy.  Of the remaining five stars, three are consistent with
being sd stars paired with a cool main sequence companion, while two may
be single stars or composite systems of another type.  We discuss the
implications of these findings for the 1125 PGR objects as a whole.  An
enlarged study is desirable to increase confidence in these first
results and to identify individual sd+cool binaries or other composites
for follow--up study.  The GALEX AIS data have sufficient sensitivity to
carry out this larger study.
\end{abstract}


\section{INTRODUCTION} \label{sec:Intro}

The subdwarf B (sdB) stars are the field analog of cluster stars lying
on the Extended Horizontal Branch (EHB), with $18000~{\rm K} \lesssim
T_{\rm eff} \lesssim 30000~{\rm K}$ and $\log g \approx 5.5 - 6$. Some
subdwarf O (sdO) stars form an extension of the EHB to higher
temperatures.  We will refer to these two groups together simply as
``hot subdwarf'' (sd) stars.  These stars may be an important source of
ultraviolet (UV) light in early--type galaxies \citep{OConnell, Brown00,
HPL07, Yi08}, and it is important to understand their origins so that
their contribution can be properly estimated.

The sd stars are core He--burning stars, with typical total mass $\sim
0.5 M_\odot$ but only very small H envelopes ($M_H \lesssim
10^{-2}~M_\odot$) \citep{Saffer}.  To become a hot subdwarf, a low--mass
star must lose its H envelope on its first ascent of the giant branch,
within 0.4 mag of the tip \citep{LateFlash}.  Since all red giants do
not end up on the EHB, some process must enhance the mass loss in stars
that do become sd stars.  Such a process could be the interaction of the
proto--sd star with a close companion star, promoting mass loss either
via Roche--lobe overflow (RLOF) or a common envelope (CE).  Such a model
was first proposed by \citet{Mengel}, and a much more detailed model has
been put forward by \citet{Han02, Han03}, including a population
synthesis study.  In the Han et al.\ scenario, the RLOF channels often
lead to a sd star paired with an A or early--F companion, the companion's
mass being increased by exchange from the proto--sd star. The CE
channels usually result in a late--F, G or K dwarf companion (or a white
dwarf companion, if the second star to evolve is the one that becomes
the sd star at the present epoch).  A binary merger channel leading to
isolated sd stars also exists.  Thus, an additional reason for
determining the evolutionary pathways that lead to sd stars is that they
may serve as a calibrating population, to enable the theory of binary
mass loss/exchange to be applied more confidently in other contexts.

We have used the 2MASS Point Source Catalog in combination with $B, V$
photometry to study the near infrared (NIR) and optical--NIR colors of a
large sample of sd stars \citep{SW, SWB}.  In a volume--limited sample,
about 25\% of these stars show an infrared excess, suggesting a
composite spectral energy distribution.  Spectroscopy of a subset of
these is consistent with the companions being mostly late--F, G, and K
dwarfs \citep{Thesis, SW06}. \citet{Lisker2005} also have found late--F,
G and K companions to hot subdwarfs, based on $B-J$ color and/or optical
spectroscopy.  These sd+G/K systems\footnote{We will use the term
``G/K'' to include late--F stars as well as G and K stars. We will use
the term ``A/F'' to refer to A and early--F stars.} are in accord with one
prediction of the binary origin scenario for sd stars.  Where are the
sd+A/F binaries that \citet{Han03} also predict? Are the existing
catalogs of identified sd stars missing a large number of sd stars in
binary systems?

With about 900 hot subdwarfs, the magnitude--limited Palomar Green (PG)
catalog of ultraviolet--excess objects \citep{PG} is the dominant
contributor to the \citet{KHD} catalog of hot subdwarf stars.  The PG
hot subdwarfs are often the subject of detailed study, either
individually or for statistical purposes. The PG candidate sample was
constructed from a $U-B$ photographic survey. Candidates showing a
\ion{Ca}{2} K--line were culled from the final PG catalog, on the
supposition that this line indicated the UV excess arose in a relatively
cool, metal--poor (Pop II) star with reduced UV line blocking (a
subdwarf F/G star$\colon$ sdF/G)\footnote{Note that despite the name,
subdwarf O/B stars are not necessarily from Pop II.  They merely appear
below the main sequence in a color--magnitude diagram.}, which was not
of interest in a survey for quasars and hot stars. Such K--line objects
with a modest $U-B$ excess, however, would also describe how an
unresolved binary composed of a hot sd star and a cool main sequence
(MS) star of type F should appear.  A relatively bright F companion
would contribute a K--line that is strong enough to be seen in the PG
survey spectra (which are of modest signal--to--noise ratio), whereas a
typical sd star would dominate over the optical spectrum and colors of a
MS G or K companion. Thus it is possible that bona--fide sd+MS binaries,
especially those with early--F companions but also late--F, G, or K
companions in some cases, are missing from the final PG catalog,
although they were selected as {\em candidates} in the PG
survey. \citet{Han03} noted this possible bias in the PG catalog as an
obstacle to confirming their binary formation scenario for hot subdwarf
stars, giving it the name ``GK selection effect''.  (Because of their
large Balmer jumps and their higher luminosities, A stars paired with
sd's would not likely have been selected in the initial PG survey for
$U-B$ excess.  On the other hand, an atypically low--luminosity sd star,
paired with a MS G or K companion, might also have a detectable K--line
in the PG survey spectra.  As noted above, G/K companions are indeed
found for many of the PG hot sd stars, based on NIR colors; thus, many
sd+G/K systems are {\em not} subject to the ``GK selection effect'', and
the term is to some degree a misnomer.)

It is desirable to have a catalog of hot subdwarf stars that is truly
representative of the space distribution of these objects. While the PG
catalog, by its design, would not be sensitive to sd+A systems, the
question has lingered among hot subdwarf aficionados whether it is less
complete than it could be with respect to sd+MS binaries where the MS
star is of type F--G--K.  Some brief quotations from recent conference
papers illustrate the concern: (1) ``The PG survey is biased against
companions of G and K spectral type as any target with a spectrum
showing \ion{Ca}{2} H--lines [sic] was taken off the survey. This should
be taken into account when looking at the companions of sdBs as most of
the ones from the PG survey will be WDs [white dwarfs] instead of main
sequence stars. This has important consequences when one intends to
compare binary formation models with observations. It is difficult to
assess how important this bias is.'' \citep{MrMM04}; (2) ``The PG
catalogue is biased against targets that show a Ca II H--line [sic] (as
these were taken out of the catalogue), and thus, against sdB binaries
with main sequence companions.'' \citep{MrMM05}; and (3) ``to avoid the
uncertain biases of the Palomar Green and other surveys...'' 
\citep{MrMM06}.

There are 1125 K--line stars rejected from the PG catalog, but retained
in an unpublished list.  If many or most of these are in fact found to
be sd+MS binaries rather than sdF interlopers as \citet{PG} supposed, 
and they turn out to have short orbital periods, the
\citet{Han03} evolution scenario (which predicts them) may be strengthened
and a missing class of sdBs found; at the same time there would be a
large increase in the total numbers of sd stars from the PG survey.
There would be important implications for the origins and also the space
density, progenitors, and formation rates of such objects. If, however,
the list of rejected PG candidates does not contain a large fraction of
composite sd+MS binaries, it is still important to know what kinds of
objects were excluded in the spectroscopic filtering of PG candidates,
so that model sd populations can be ``observed" in the same way that the
PG catalog was constructed.  Which is the actual situation?

Optical studies (broad--band photometry, moderate resolution
spectroscopy) cannot easily distinguish an sdF star from a sdB+F
composite.  Metal--poor sdF stars lie above the Population I main
sequence in the $U-B$, $B-V$ two--color diagram, owing to reduced line
blocking, and thus may possibly be confused with composite sd+F systems,
where the UV excess comes from the sd's contribution. Likewise, an sd
star would dilute the metal lines in a Population I F star's spectrum
while maintaining the strength of the hydrogen lines (since these appear
in both spectra).  Similar arguments apply to the confusion between sdG
and sd+G/K systems, if the sd star has a lower luminosity; such low
luminosity hot subdwarfs are predicted by \citet{Han02, Han03} to come
about via evolution from intermediate mass stars which undergo RLOF mass
loss.  In the far UV, however, the distinction between sdF/sdG stars and
sdB+MS composites is easy to make. With its UV sensitivity, positional
accuracy and precision, and wide sky coverage, the GALEX satellite
provides an opportunity to determine the nature of the stars rejected
from the PG survey.

We note that the combination of far UV photometry from space with
ground--based work has led in the past to the identification of sd+MS
systems. An early example is the case of HD 17576, studied with the
S2/68 experiment about the TD--1 satellite \citep{Darius78, Olsen80}.
Although this system is actually a visual binary with an angular
separation of 1\farcs8, the photometry was of the blended sdO+G pair.
Another example is that of HD 15351, a sd+F5V system \citep{Darius84}.
Had these been fainter systems, they might have been selected as
candidates in the PG survey, then rejected owing to the presence of a
K--line in the optical spectrum.

To summarize, the leading scenario for the formation of sdB stars
predicts that many sd+MS systems should exist, constituting a large
fraction of all hot subdwarfs, but sd+A/F systems are hard to identify
from optical studies.  The K--line stars rejected from the PG survey are
the obvious and first place to look for some of these objects. A study
using GALEX photometry for a subset of the PG rejects, combined with
existing visible and NIR information on these stars, allows a clear
sorting into mutually exclusive categories of metal--poor sdF/G stars and
the more interesting sd+MS cases.  If these objects are in fact sd+MS
binaries, they will be strongly detected in both the near--ultraviolet
(NUV) and far--ultraviolet (FUV) GALEX passbands.  If they are actually
sdF/G stars, as originally supposed by the compilers of the PG catalog,
then they will be readily detected in the GALEX NUV band but {\em not}
in the FUV band.  Such a study addresses the origin of hot subdwarf
stars very directly. Further, if the yield of sd+MS binaries from these
rejected PG candidates is high, it may help to create an enlarged and
more representative catalog of such objects.

In this paper, we present a first assessment of the sd+MS binary content
among the PG--rejected stars.  In \S2 we present the far UV, visible, and
NIR photometry for a sample of 88 objects. We also present results for a
small number of known hot subdwarfs and other stars, and for field F
stars, to facilitate interpretation of the results in terms of modeled
colors.  We interpret the results in \S3, where we discuss five
individual objects that can be interpreted as composite hot+cool
objects, but only three of which are likely consistent with being sd+MS
binaries. We also discuss the implications for the entire group of
1125 PG--rejected stars as a whole, and we point out the desirability of
an enlarged study, both to increase confidence in this preliminary
results and to identify individual sd+MS or other composite objects for
follow--up study.  We summarize our findings in \S4.


\section{PHOTOMETRY OF PG--REJECTS} \label{sec:Data}

The essential idea underlying this study is to test, for each
``PG--Reject'' (PGR) object, whether the UV energy distribution indicates
the presence of a hot stellar source accompanying the star that is
responsible for the K line or the G band.  If no hot subdwarf is
present, the far UV flux will fall steeply toward shorter wavelength,
and an ultraviolet--based color will be very ``red''. In this case the
favored interpretation is that the PGR is a single, late--type MS star,
possibly metal--poor owing to its method of selection. (We use
solar--metallicity stars to model the composite systems, since we expect
the supposed hot subdwarf in such a system to have formed from an
intermediate mass star.) We use the GALEX archive to collect the UV flux
information.  The GALEX spacecraft is currently performing a number of
imaging surveys over most of the sky, using a far--ultraviolet ``FUV''
band (1350--1750 \AA) and a near--ultraviolet ``NUV'' band (1750--2750
\AA). See \citet{Martin05} for a general description of GALEX and
\citet{Morrissey07} for details of the instrument calibration and data
products.  We also make use of NIR photometry to aid in classifying the
cool stars in these systems (\S\ref{NIR}).

We use $F$ and $N$ in formulas as symbols for the AB magnitudes
measured in the FUV and NUV bands, respectively.  The $F-N$ color is
the main quantity of interest, but interpretation is enhanced with the
addition of the $N-V$ color, where $V$ is from the Johnson $UBV$
system.  We employ synthetic photometry as a guide to the expected
colors of single or blended objects.

From the 1125 PGRs, we selected 88 as the subject of this first
assessment using GALEX.  
These 88 are the majority of the PGRs for which GALEX photometry
was available in data release GR1.
Table~\ref{table1} presents the names of these
objects along with the names of their 2MASS
\citep{Skrutskie06} counterparts.  The IAU--registered names are of the
form {\tt PGR JHHMMm+DDMM}, where minutes of Right Ascension (RA) are
truncated after the first decimal. The designation style is distinct
from that used for PG objects, and is sufficiently detailed to
give a unique name to each PGR object.  These names are based on the
coordinates of the objects as recovered from the USNO--A2 catalog
\citep{usno}, which is based on the Palomar Observatory Sky Survey
(POSS) plates taken in the 1950s.  (The original finding charts for
the PGRs are enlarged prints from the POSS.) The 2MASS names are also
in equatorial coordinate form, based on observations at a much more
recent epoch. The 2MASS names suffice to allow precise and unambiguous
identification of the star, thus we do not provide separate columns with
RA and Declination.

\placetable{table1}
\notetoeditor{Table 1 (long version) is to be electronic only;
that file is called 'table1-long.tex'.  A portion of the table is to be
shown in the printed version for guidance; the file for this version is
called 'table1-short.tex'. We have used the LaTeX label 'table1' for
BOTH tables.  In the main file 'ms.tex' as submitted, we call the long
version.}

\subsection{GALEX Photometry}

Ultraviolet photometry of the PGRs as obtained by GALEX is presented in
Table~\ref{table2}, along with an estimate of the $V$ magnitude (see
\S\ref{sec:Visual}), color excess E$(B-V)$, and colors derived from the
photometry.  GALEX Release 2/3 was most recently interrogated in May
2008, using the Cross--Correlation search page of the Multimission
Archive at Space Telescope (MAST)\footnote{{\tt
http://archive.stsci.edu/index.html}}.  The search radius used was 6
arcsec.

Table~\ref{table2} lists $F$  and its error in the AB system
\citep{Morrissey07} or an upper limit on $F$.
It also lists $N$ and its error.  All the PGRs discussed
here were detected by GALEX in the NUV band.  Sixteen were also
detected in the FUV band. In the case of multiple observations of a
source by GALEX, we merged the data into single best estimates of $F$
and $N$ using weighted averages.  The quoted errors refer to these
best estimates.  Magnitudes and colors are rounded to two
decimal places, with a floor of 0.01 mag placed under the magnitude
errors.  When a source was not detected in the FUV band, we calculated
the flux upper limit as three times {\tt fuv\_ncat\_fluxerr}, which is the
tabulated error in the net FUV flux, measured at the position of the
NUV detection.  This flux limit is translated to AB magnitude units;
the $\sigma(F)$ error column in Table~\ref{table2} is left blank for upper
limits.  

\placetable{table2}
\notetoeditor{Table 2 (long version) is to be electronic only;
that file is called 'table2-long.tex'.  A portion of the table is to be
shown in the printed version for guidance; the file for this version is
called 'table2-short.tex'. We have used the LaTeX label 'table2' for
BOTH tables.  In the main file 'ms.tex' as submitted, we call the long
version.}

Of the sixteen FUV detections, twelve were from short GALEX
observations that were part of the All--sky Imaging Survey (AIS), one
was from a longer Medium Imaging Survey (MIS) observation, two were
from longer Guest Investigator (GI) observations, and one was from a
very long observation in the direction of the Lockman hole (LOCK).
FUV--band non--detections of the other 72 PGRs were generally for
sources only observed in AIS fields, but in a few cases MIS, GI, LOCK,
or Nearby Galaxy Atlas (NGA) observations were able to set more
stringent upper limits on the FUV flux.  Remarks in Table~\ref{table2}
indicate whether observations other than AIS were available.

\subsection{Visual Photometry}\label{sec:Visual}

Estimates of $V$ are generally only photographic, since the PGRs are
generally too faint to have been the subject of individual
photoelectric photometry and too bright to be observed without
saturation by the Sloan Digital Sky Survey (SDSS) in the area of
overlap.  For the photographic $V$ estimates, we used the prescription
from \citet{SalimGould03} to convert from the USNO--A2 red and blue
magnitudes: $$ V = R_{\rm USNO} +0.32(B_{\rm USNO} - R_{\rm USNO}) +
0.23$$ These are indicated in Table~\ref{table2} by the symbol
``A'' in the ``Ref'' column.  The symbol ``B'' in this column
indicates that $V$ is estimated from SDSS $g$ and $r$ magnitudes
according to the prescription of \citet{Jester05}: $$ V = g -
0.58(g-r) - 0.01$$ 
We only used SDSS data that were unsaturated in both $g$ and $r$.
Symbol ``C'' indicates photoelectric $UBV$
photometry carried out by R.F.G.\ during the original PG survey.
Symbol ``D'' indicates photoelectric CCD photometry in $V$ and $I$ by
the All Sky Automated Survey\footnote{{\tt
http://www.astrouw.edu.pl/asas/}}
\citep[ASAS:][]{Pojmanski02}. 
In total, 24 PGRs in Table~\ref{table2} have photoelectric photometry.
The tabulated $V$ magnitudes are rounded to 0.1 mag. 
Based on the references cited or our own investigations, the
uncertainties in the transformed or measured $V$ magnitudes are
$\sigma(V) \approx 0.03$ for objects observed photoelectrically by SDSS;
$\sigma(V) \approx 0.07$ mag for R.F.G.; and
$\sigma(V) \approx 0.05$ for ASAS.
Photographic estimates have $\sigma(V) \approx 0.25$ mag.

Colors $F-N$ and $N-V$ are given in Table~\ref{table2}, along with a
dereddened color $(N-V)_0$, where we have taken the color excess to be
$$E(N-V) = 4.8 E(B-V)$$ as suggested by the GALEX exposure time
calculator for a flat--spectrum source ($f_\nu = {\rm const}$) and
assuming that $A(V) = 3.2 E(B-V)$.  The interstellar reddening of
$F-N$ is small, $E(F-N) \approx -0.1 E(B-V)$, and this correction has
been neglected.  The tabulated $E(B-V)$ estimates are those returned
from the MAST query, based on the \citet{Schlegel} reddening
maps. They represent the full galactic extinction along the line of
sight and should be appropriate for these PGR objects, which are
generally expected to lie well above the Galactic plane.  Color errors
can be estimated by combining in quadrature the errors of the
individual bands.

The $V$ magnitude range is 11.2 to 16.3 (median near 14.7, although
for PGRs with detected flux in the FUV band the median $V$ is 13.3).
The range of $N$ magnitudes is 13.7 to 20.4.  {\em Detected} $F$
magnitudes range from 13.5 to 24.3; {\em upper limits} on $F$ range
from 21.1 to 23.5.  Color excess $E(B-V)$ from \citet{Schlegel} ranges
from 0.008 to 0.264 mag.

\begin{figure}
\epsscale{0.70}
{\plotone{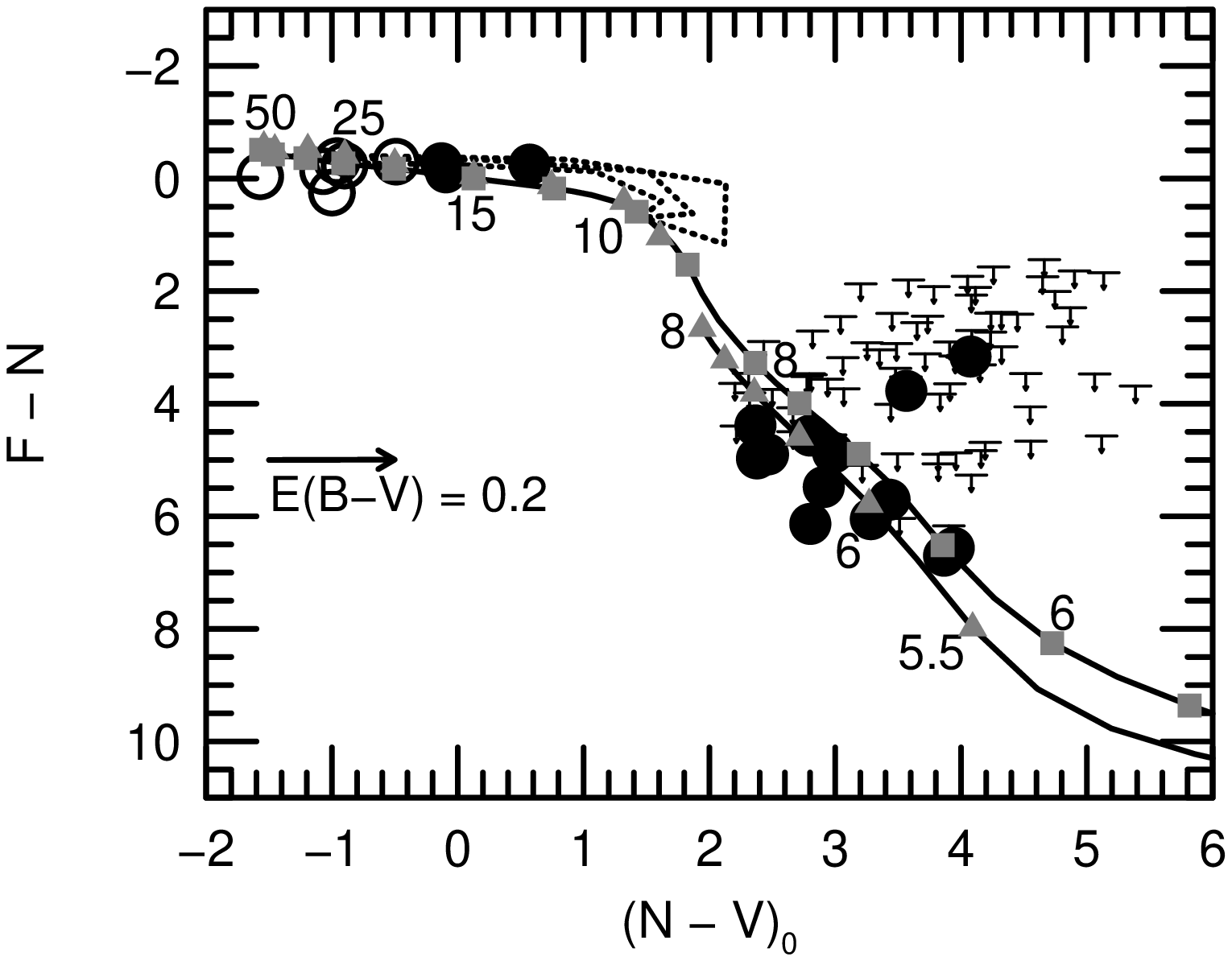}}
\caption{Observations of PGR objects, presented in a two--color diagram
using Johnson $V$ and GALEX $F$ and $N$ magnitudes.
Filled circles are PGR objects detected in both $F$ and $N$;
horizontal bars are PGR objects detected in $N$ but with upper
limits in $F$ (Table~\ref{table2}).  
Open circles are known hot stars (Table~\ref{table3}).
A reddening vector for Milky Way dust is shown. 
Solid lines show model loci for solar and metal--poor synthetic spectra
at $\log g = 5.0$ \protect{\citep{Kurucz}}.
Grey squares ([Fe/H]$=0.0$) and triangles ([Fe/H]$=-1.5$) along the loci 
are shown for $T_{\rm eff}$ = 5.5 (0.5) 8.0 (1.0) 10, 12, 
15 (5) 30 (10) 50 kK, with selected points labeled.  
Dotted lines show colors of composite models, combining a ``hot
subdwarf'' and a cool main sequence star (see \S\ref{sec:Visual}).
\label{fig1}}
\end{figure}

Figure~\ref{fig1} presents a $(F-N)$, $(N-V)_0$ two--color plot of the
observations. Solid lines show model loci for solar and
metal--poor synthetic spectra at $\log g = 5.0$
\citep{Kurucz}, convolved with the GALEX FUV
and NUV bandpasses and an approximate Johnson $V$ bandpass. The GALEX
post--launch response curves were obtained from the GALEX Guest
Investigator website. The $V$ bandpass was approximated as having a
square response over 5150--5950 \AA.  The metal--poor locus is truncated
at 8000 K, somewhat hotter than the Pop II turnoff.  Dotted lines show
the colors of composite models, combining a sd and a cool MS star to
make a sd+MS system.  For these loci, the subdwarfs are represented by
models having [Fe/H] = 0.0, $\log g = 5.0$.  These models are assigned
$M_V = 3.59, 4.05$, and $4.60$ for $T_{\rm eff} = 25000, 30000$, or
$35000$ K, respectively, derived from the luminosities of zero--age
extended horizontal branch (ZAEHB) models of
\citet{Caloi} using suitable bolometric corrections. Each ZAEHB
model generates one of the three loci shown, as it is combined with
light from the companion MS star.  The MS stars are represented by
models having [Fe/H] = 0.0, $\log g = 5.0$ and $T_{\rm eff}$ ranging
from 4000 K to 9750 K. Absolute $V$ magnitudes of the cool models are
adopted MS values.  The loci of composite models ``loop back'' to the
left in the Figure, as the MS component's $N-V$ color index decreases
with increasing $T_{\rm eff}$ at the same time that the MS component
begins to dominate the combined light of the sd+MS system.  For the
adopted ZAEHB absolute magnitudes, this happens when the MS star has
an F spectral type.

\placefigure{fig1}

Open circles in Figure~\ref{fig1} show the colors of six stars
classified by \citet{PG} as hot subdwarf stars
(Table~\ref{table3}). Estimated reddening is again from
\citet{Schlegel} via MAST; here we have dereddened both $F-N$ and
$N-V$ to facilitate comparison with the model loci. Note, however,
that non--linearity of the GALEX photometry at these magnitudes may be
important \citep{Morrissey07}.  From 2MASS photometry
\citep[see, e.g.,][]{SW}, PG 0105+276 and TON 349 are regarded as
photometrically ``composite" while the remaining stars are regarded as
single.  Note that PG 0105+276 is a close visual pair with separation
$\approx 4''$; we treat the $F$, $N$, and $V$ measurements for this
object as referring to the combined light from this pair.

\placetable{table3}

\subsection{Near--Infrared Photometry}\label{NIR}

\begin{figure}
\epsscale{0.7}
{\plotone{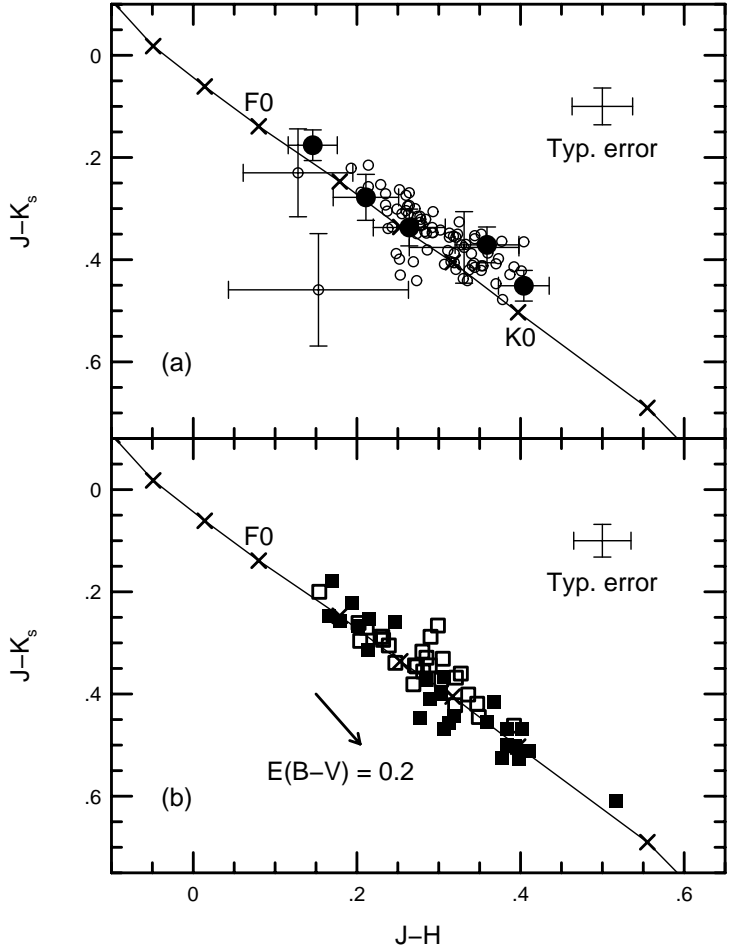}}
\caption{(a) NIR two--color diagram showing the PGR objects.
The five ``hot'' or ``comp?'' objects of Table~\protect{\ref{table4}}
are shown as filled circles; other PGRs are shown as small open circles.
Individual error bars are shown for these five objects
or in cases where $\sigma(J-H)$ or $\sigma(J-K_s)$ exceeds 0.07 mag.
The median (``typical'') error bar is shown in the upper right.
The MS locus of \protect{\citet{B+B}} is shown, converted
to 2MASS colors using the prescription of \protect{\citet{Carpenter}}; 
crosses indicate spectral types of A0, A5, F0, F5, G0, G6, K0, and K5.
(b) The same diagram, showing the colors of the ``Cayrel'' sample 
(see \S\ref{NIR}).
Filled squares have ``solar'' metallicity; open squares are ``metal--poor''.
A reddening vector for \protect{$E(B-V)=0.2$} is shown.
No reddening correction has been applied to data in either panel.
\label{fig2}}
\end{figure}

NIR colors from 2MASS for all of the PGR stars in this study
are shown in panel (a) of Figure~\ref{fig2}.  The MS locus of
\citet{B+B} is also shown, converted to 2MASS colors using the
prescription of \citet{Carpenter}.  Panel (b) of Figure~\ref{fig2}
shows the NIR colors of stars selected from the abundance compilation
of \citet{Cayrel+01}, with $V > 8.5$, $T_{\rm eff} > 5000$~K, $\log g
> 3.5$, and $E(B-V) < 0.20$.  We retained 27 Cayrel stars with [Fe/H]
between $-0.3$ and $+0.3$ (``solar'') and 27 stars with [Fe/H] between
$-1.8$ and $-1.2$ (``metal--poor'').  These two groups are
indistinguishable by location in the NIR two--color diagram; moreover,
the NIR colors of PGR objects closely follow the same locus within
the errors of measurement.

\placefigure{fig2}


\section{DISCUSSION} \label{sec:Discuss}

\subsection{Overview}

Inspection of Fig.~\ref{fig1} shows that among the PGRs with
detections in the FUV band, three lie near the hot end of the stellar
locus (``hot'' is noted in the Remarks column of Table~\ref{table2}
for these). Eleven lie near the loci shown for single stars at $T_{\rm
eff}\sim 7000$~K, and two lie above and to the right of this position
(``comp?'' in Table~\ref{table2}).  We summarize the GALEX, visual and
NIR photometry for the five ``hot'' or ``comp?'' stars in
Table~\ref{table4}.  All of the upper limits on $F$ result in
placement of PGRs well away from the part of the diagram that
characterizes blended light from a sd+MS system.  This is the main
result of our study: {\em most of the PGRs in this sample are not
binary systems containing a hot subdwarf star and a MS star}.

\placetable{table4}

\subsection{Validity of GALEX Photometry}

The three ``hot'' PGRs all have $F$ and $N$
magnitudes brighter than 15.4.  The measured count rates in the GALEX
detectors for these brightness levels may be subject to non--linearities
and other ``saturation'' or ``fatigue'' effects, characterized {\em
statistically} by ``roll--off curves'' \citep[see][Fig.~8]{Morrissey07}.
For the brightest of the three ``hot'' PGRs, PGR J00075+0542, MAST
returned two distinct entries, with $F$ differing by 1.0 mag and $N$
differing by 0.3 mag in the opposite sense.  Clearly, the vertical
position of this object in Fig.~\ref{fig1} is uncertain by $\sim$0.5
mag, much larger than the formal error bars reported by MAST and
tabulated here. Fortunately for this study, even an error of 0.5 mag in
$F-N$ does not vitiate the conclusion as to the nature of this object,
which clearly contains a hot star.  (As we did for all other stars with
multiple observations, we have averaged the magnitudes for PGR
J00075+0542.)

Some of the scatter in $F-N$ (and $N-V$) for the known hot subdwarfs of
Table~\ref{table3} may also be due to the ``non--linearity'' and other
performance issues in GALEX photometry of bright point sources, as
discussed above.  In this paper, we take the reported photometry and
errors at face value for discussion purposes, since it does not affect
the bulk of our sample.  Detailed modeling of bright composite objects
should not rely too heavily on GALEX photometry, however.

\subsection{Validity of Model Loci}\label{model-validate}

Despite the uncertainty in positioning the known hot subdwarfs in
Fig.~\ref{fig1}, it is evident that the synthetic photometry is in
reasonable agreement with the observations of these hot stars.  PG
0105+276 has been classified as sdO(B) and sdB+K7 (visual double).  PG
0212+148 has $T_{\rm eff}$ near 25000~K.  PG 2356+167 is classified
sdB-O by \citet{PG} but ``non--sd'' by \citet{Saffer1991} who notes the
\ion{Ca}{2} K line in the spectrum; \citet{Saffer1997} estimate 
$T_{\rm eff} = 23800$~K, $\log g = 4.70$, while \citet{Lynn2004}
classify this object as B2V (evolved) and estimate $T_{\rm eff}\approx
20000$~K and $\log g = 4.3$.  The remaining stars are classified
``sdO(B)'' or ``sdB'' by \citet{PG}.

\begin{figure}
\epsscale{0.70}
{\plotone{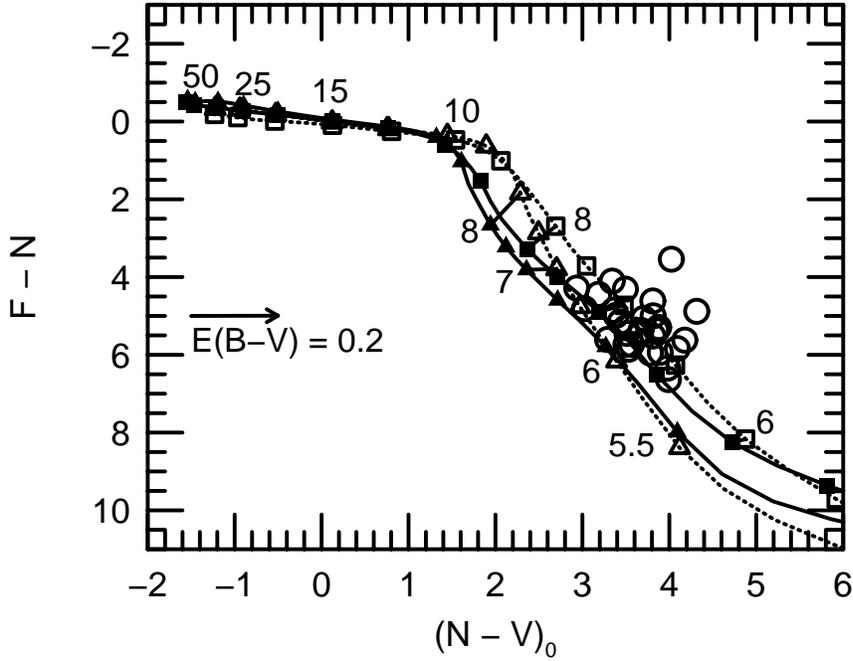}}
\caption{Observations of candidate F dwarfs, selected on the basis of
their 2MASS colors, presented in the same two--color diagram as shown in
Fig.~\protect{\ref{fig1}}.  The observed $N - V$ colors have been
corrected for galactic reddening.  Model loci at $\log g = 5.0$ (solid
lines) are shown and labeled as in Figure~\ref{fig1}, supplemented by
loci at $\log g = 3.5$ (dotted lines).  All loci extend to $T_{\rm eff}
= 30000$~K, with the $\log g = 5.0$ loci extending to 50000 K. Filled
squares and triangles show $\log g = 5.0$ models at metallicity [Fe/H] =
0.0 and $-1.5$, respectively; open symbols show $\log g = 3.5$
models. ``Tie--bars'' connect low-- and high--gravity model points for each
metallicity at $T_{\rm eff} = 6000$, 7000, and 8000~K.
\label{fig3}}
\end{figure}

To see how well the model loci of cooler single atmospheres match the
GALEX observations, we collected photometry for a sample of 30
lightly--reddened F star candidates ($E(B-V) < 0.04$), selected on the
basis of their 2MASS colors. The intrinsic 2MASS colors of F0--F7 dwarfs
were determined from stars classified on the MK system by
\citet{HoukVol5}.  $V$ magnitudes for the sample discussed here
were collected from ASAS and range between 11.2 and 12.5. The GALEX $N$
magnitudes range from 14.6 to 17.0 with nominal errors $\sigma(N) <
0.03$, while $F$ ranges from 18.9 to 23.2 with median error $\sigma(F) =
0.23$ (maximum $\sigma(F) = 0.52$).  These stars are displayed in the
$(F-N)$, $(N-V)_0$ two--color diagram in Figure~\ref{fig3}, where typical
$1\sigma$ error bars are the size of the plotting symbol or smaller.
The model loci for single stars from Figure~\ref{fig1} are supplemented
with additional loci.  

\placefigure{fig3}

The observed candidate F stars generally lie near the [Fe/H] = 0.0 loci
as expected, but there is some scatter to ``bluer'' $F-N$ values,
somewhat larger than might be expected from the typical error of
measurement. (The range of $T_{\rm eff}$ inferred from the $(N-V)_0$
colors of these stars is in good accord with that of F2--F7 dwarfs of
Population I, as suggested by the 2MASS photometry.  Thus, the scatter
does not seem attributable to $(N-V)_0$.)  A tentative interpretation is
that the upward scatter may be caused by unmodeled chromospheric or
transition region activity which varies in strength from star to
star. This might be manifested, for example, by emission at the
\ion{C}{4} 1548,\,1550 \AA\ doublet, which is located near the peak of
the GALEX FUV passband.  With this caveat, we conclude that the single
star synthetic photometry based on \citet{Kurucz} models provides an
acceptable basis for constructing colors of composite models and for
interpreting the photometry of PGR objects.

\subsection{Interpretation of the Two--Color Diagrams}

\citet{SW} showed that hot subdwarf stars with $J-K_s
\gtrsim +0.15$ form a distinct group which can be modeled as
photometrically composite, with the cool companion star consistent
with a dwarf of spectral type (SpT) F, G, or K.  This result is
confirmed by spectra of a subset of such systems
\citep{Thesis, SW06}.  Of course, late--type stars that are single also
have red $J-K_s$ colors, thus 2MASS data alone do not suffice to
indicate whether a red object is a composite system with a hot
subdwarf and a cool star.  It is the combination of GALEX and 2MASS
data that allows a composite system to be recognized.

Based on the GALEX photometry shown in Figure~\ref{fig1}, there is a
hot star present in each of the three PGRs labeled ``hot'' in
Table~\ref{table4}.  Considering also their NIR and $V-K_s$ colors,
they are all photometrically composite.  If PGR J00075+0542 were
a single star, its NIR colors would indicate a SpT near F2, but the
cool star may be slightly later than this, given the dilution by the
hot component.  Dilution should be more evident in the $V-K_s$ color
index, and indeed it suggests a SpT near F0 under a single--star
interpretation. By the same reasoning PGR J22451+2134 contains a
cool star with SpT somewhat later than K0.  PGR J23025+2602's
cool component must have SpT later than about F8.  The GALEX
photometry supports this ordering by color (given the uncertainty in
the measured $N$ magnitudes).  In particular, PGR J00075+0542 has the
earliest cool component based on visual and NIR colors, and has the
{\em reddest} $(N-V)_0$ color among the three ``hot'' PGRs, a result
entirely consistent with combining some hot ZAEHB subdwarf with MS
companions of different temperatures and luminosities in the three
cases.  

\citet{Schuster04} independently studied PGR J00075+0542, their name
for this object being BPS CS 31070--0080.  They classified it as a
``sub--luminous blue horizontal branch star'' (SL--BHB).

The $(N-V)_0$ indices are consistent with roughly equal absolute $V$
magnitudes of the hot and cool components in each system.  The
interpretation that these three PGRs are sd+MS systems is perhaps not
unique; however, the spectroscopic type of ``sdG'' assigned by
\citet{PG} in each case supports the argument that the cool components
are dwarfs rather than evolved stars.  Pending further study to establish
more accurate temperatures, gravities, and metallicities, we adopt the
sd+MS picture as the best interpretation of the data.

Based on GALEX and $V$ data alone, the two ``comp?'' stars in
Table~\ref{table4} might simply be F or G dwarfs with high activity
levels: they lie within the extreme limits of the scattering of
candidate F stars in Figure~\ref{fig3}. On the other hand, a composite
system is not ruled out in either case, although it would not be
consistent with a MS star paired with a ZAEHB star.  The NIR colors
for PGR J08401+4421 correspond to a (single) dwarf of SpT near
G0, while $V-K_s$ suggests the SpT is near G5 and $(N-V)_0$ suggests
$T_{\rm eff}\approx 6800$~K.  These are slightly inconsistent with the
expectation for a hot+cool binary, in terms of the ordering of implied
temperature by wavelength.  Nevertheless, we conducted numerical
experiments, adding a hot component to a Kurucz model with $T_{\rm
eff} \approx 6800$~K. The results suggest that such a hot component
would need to be about 4 mag fainter than the cool star at $V$, in
order to reproduce the GALEX colors.  In such a case, the hot star
would be fainter than a typical hot subdwarf, or the cool star would
need to be brighter than a MS star.  A white dwarf is a possible hot
component, although $T_{\rm eff} \gtrsim 30000$~K is required if its
radius is that of a $\sim 0.6\,M_\odot$ remnant.  The cooling time of
such hot white dwarfs is short \citep{WoodWinget89}.  A similar
analysis applies for PGR J02040+1500, whose NIR and $V-K_s$
colors suggest a SpT near G5 or G8, but whose $(N-V)_0$ index suggests
a warmer star with $T_{\rm eff}\approx 6500$~K (using the solar
metallicity loci).  The $F-N$ and $(N-V)_0$ colors can be roughly
matched by combining a $T_{\rm eff} \sim 6500$~K star and a hot star
that is $\sim$4 mag fainter at $V$.  Improved visual photometry of
these systems and accurate luminosity (and metallicity) classifications
of their cool stars would help clarify the nature of the hot
component.

The remaining eleven PGR objects that were detected in the FUV band
lie close to the low--metallicity (Pop II) MS locus in
Figure~\ref{fig1}, although with some scatter.  This is consistent
with the spectroscopic classification of ``sdG'' (with slight
variants) given to these objects in the original PG survey.  No
indication of the presence of a hot stellar component is given by the
GALEX photometry of these stars.  The 72 upper limits for other PGRs
shown in the Figure are sufficiently far removed from the locus of
composite sd+MS loci that they likewise support the original sdG
classifications of these objects as single stars.  The NIR colors of
the ensemble are consistent with a single--star interpretation,
although it should be noted that the (Population I) SpTs indicated
extend to K0 ($T_{\rm eff} \approx 5200$~K), whereas the $(N-V)_0$
data suggest a lower bound of $T_{\rm eff} \gtrsim 5700$~K (solar
metallicity) or $T_{\rm eff} \gtrsim 5200$~K (metal--poor).

Three stars, PGR J09281+6503, PGR J09395+6353, and PGR J10273+5758,
for which $UBV$ data are available, lie slightly above the Hyades main
sequence in the $(U-B)$, $(B-V)$ two--color diagram with $B-V$ in the
range $+0.50$ to $+0.63$ and $U-B$ in the range $-0.124$ to $+0.07$,
just where reduced metal line blanketing would place a sdG star.  The
first of these stars was detected in the FUV band and lies very near
the low--metallicity locus in Figure~\ref{fig1}, near $T_{\rm eff} =
6000$~K.

The occasional large offsets between USNO--A2 and 2MASS catalog
positions, noted in Table~\ref{table1}, suggest that a few stars of high
proper motion may be included in this sample of PGR objects.  This view
is confirmed by proper motion data for 61 stars of the sample, obtained
from the Second U.S.\ Naval Observatory CCD Astrograph Catalog
\citep[UCAC2:][]{Zacharias2003}.  Stars with large Table~\ref{table1}
offsets tend to have larger than average components of proper motion to
the West, with amplitudes $\mu_\alpha \sim -25\,{\rm mas~yr^{-1}}$
(corresponding to transverse velocities $v_T\sim 100\,{\rm km~^{-1}}$ for
distances $d\sim 1000$~pc).  The direction of motion is opposite to that
of the Local Standard of Rest, and thus an interpretation in terms of
thick disk or galactic halo rotation for this group of stars is
tempting, although only weakly indicated.

\subsection{Implications for the PGRs as a Class}

The PG catalog is the single largest contributor to the list of
spectroscopically identified hot subdwarf stars \citep{KHD}.  Because it
has reasonably well--defined magnitude and color limits, it is often used
as the source of moderately bright sdB/sdO stars on which to do
time--series photometry, radial velocity orbital studies, etc.  It also
has been used for comparison against population synthesis models in
studies of the origin of hot subdwarfs.  Every catalog that strives to
be complete will suffer from two types of error, that of including
objects that do not belong, and that of excluding objects that do
belong.  Concerns have been raised about the hot subdwarf part of the PG
catalog, in regard to this second type of error. For this reason we
looked again at a sample of the stars rejected from the PG catalog.

Only three of the 88 PGR objects examined in this paper show a
photometrically composite character, consistent with a sd+MS system.
Two additional objects {\em may} be photometrically composite, although
in these cases the data do not support a sd+MS interpretation.  The
great majority of the PGR stars appear to be consistent with single
stars on the lower MS, perhaps chromospherically active, perhaps metal
poor.  Additional data, such as $UBV$ colors for a few stars, support a
metal--poor classification in those cases.  As originally explained by
\citet{PG}, the PG survey photographic photometry was imprecise enough
to allow the selection of numerous {\em candidate} UV--excess stars which
were in fact not ``hot'' in the desired sense of meeting a color
threshold of $U-B < -0.46$.  The survey spectroscopy was used to weed
these out, leaving only ``hot'' objects in the final PG {\em catalog}.
That the rejected candidates (PGRs) are almost entirely metal--weak sdF
or sdG stars is entirely consistent with the nature of the PG survey;
this is because the most likely contaminant, based on a measured $U-B$
color accidentally crossing the color threshold for the survey, would be
an F or G star with reduced metal line blanketing.  That a few systems
were rejected which do in fact contain a hot star (likely a subdwarf B
or subdwarf O star) should not be especially surprising, if they were
paired with cooler stars, depending on which star dominates the
photographic region of each blended spectrum.

This paper exploits the wider wavelength range enabled by the GALEX and
2MASS projects to further characterize these systems. The quantitative
result is that the fraction of such hot+cool systems among the PGR
objects that comprise the present sample is small.  An important
question is whether this conclusion extends to the entire list of 1125
PGRs.  The PGR list is not uniform with respect to $U-B$, nor with
respect to limiting $B$ magnitude.  This is because in some PG survey
fields, classification spectroscopy was started before a final $U-B$
color (transformed from photographic magnitudes) was available. Thus
some UV--excess {\em candidates} were rejected that were ultimately
measured to be redder than the color cutoff of the PG survey.  The
present sample was drawn from a subset of PG survey fields that
were observed early in the GALEX observational program, and it may draw
more or less heavily from PG fields with an ``extra'' supply of PGRs.  A
wider study of the PGR list is therefore needed to confirm whether the
present result, a $\sim 3/88$ rate of incidence of sd+MS systems, is
representative of the PGR stars overall. Because of the non--uniformity
of the PGR list, this would be not merely a refinement in the
statistical precision of the present result, but a necessary
confirmation or improvement. An additional benefit of a significantly
larger sample would be to more firmly establish the ratio of F--type
companions to G-- or K--type companions in the sd+MS systems that are
found.  In the \citet{Han02, Han03} binary population synthesis models,
this ratio is strongly related to the critical mass--ratio parameter,
which serves to separate stable and unstable mass transfer cases.
Finally, it is important to identify those individual PGR stars from the
entire list that {\em are} photometrically composite, for appropriate
follow--up studies, including the measurement of orbital periods.

We can attempt a cautious extrapolation of the results of the present
sample of 88 stars to the full 1125 PGR objects.  At present, there is
no evidence to suggest that the present sample of 88 differs
systematically from the full sample, so such extrapolation may be a
useful guide to what may be expected, bearing in mind the caveats stated
above. Taking 3/88 as the rate of incidence of sd+MS systems in the full
sample, we predict ~$\sim$38 ``new" sd systems will be found among the
PGR stars. This is to be compared with the $\sim$40\% of sdB stars in
the \citet{KHD} catalogue that were found by \citet{SWB} to have a cool
companion; applied to the $\sim$900 PG stars in \citet{KHD}, this
amounts to $\sim$360 sd+MS stars.  The ``new" sd+MS systems that might
result from a study of the full PGR list thus represent a $\sim$10\%
increase in this roughly magnitude-limited sample.  We would not be
surprised to find that the rate of incidence of sd+MS systems in the
full PGR list may be $\sim 2\times$ smaller or larger than 3/88 (3.5\%),
either from small-number statistics or ``bad luck" in drawing our small
sample from a few fields.  Thus the addition of new hot sd+MS systems
from among the PGR objects to those found in the PG catalog may
plausibly be expected to be at the 5--20\% level, a consequence of the
large numbers of PGR objects, coupled with a fairly small rate of
incidence of sd+MS binaries in the PGR list.


\section{SUMMARY} \label{sec:Conclude}

We have presented UV, visual, and NIR photometry of a sample of
candidate objects rejected from the final PG catalog. We analyzed the
photometry in color--color diagrams, aided by synthetic photometry of
single and binary stars and by empirically determined colors of stars of
various types.  Sixteen of the 88 PGRs were detected in both the FUV and
NUV channels of GALEX imaging.  Of these, eleven are consistent with
being single cool stars, as are the 72 cases where only an upper limit
on the FUV flux was obtained.  Thus, most of these PGRs have colors that
are consistent with those of single cool stars, possibly metal--poor.
This is consistent with the removal of these candidate UVX stars from
the PG survey, based on spectroscopy.  Of the remaining five stars,
three exhibit composite colors consistent with a sd+MS interpretation,
and the other two may be single or composite.

The GALEX AIS survey is sufficient to detect a hot component in a PGR,
if one is present. Thus it should be possible to extend the present
study of 88 PGRs to include most of the 1125 objects in the full list,
which is desirable given the non--uniform nature of the PGR list.  Such
an extension would establish definitively whether the PGR list does or
does not harbor a large number of sd+MS binaries or other hot+cool
systems, putting to rest the concern about the completeness of the PG
catalog (within its declared magnitude and color limits) with regard to
hot subdwarfs, hot white dwarfs, etc. It would provide a more robust
measurement of the frequency of occurrence of early--F companions,
compared with late--F/G/K companions. Finally, it would identify
those individual PGRs that are photometrically composite, to allow
appropriate follow--up studies. We are currently undertaking this larger
study.

\acknowledgments

We thank David Monet for providing a copy of the USNO--A2 catalog, and
John Feldmeier and Bruce Koehn for assistance in the installation and
use of the software package Refnet, which we used to access the catalog.
We also thank Luciana Bianchi for helpful correspondence regarding
synthetic GALEX  photometry.
We thank the referee for helpful and constructive comments.
Support from NASA grants NGT5--50399, NNG05GE11G,
and NNX09AC83G is gratefully acknowledged.
Based on observations made with the NASA Galaxy Evolution Explorer.
GALEX is operated for NASA by the California Institute of Technology
under NASA contract NAS5-98034.
This publication makes use of data products from the Two Micron All Sky
Survey, which is a joint project of the University of Massachusetts and
the Infrared Processing and Analysis Center/California Institute of
Technology, funded by the National Aeronautics and Space Administration
and the National Science Foundation. 
Some of the data presented in this paper were obtained from the
Multimission Archive at the Space Telescope Science Institute
(MAST). STScI is operated by the Association of Universities for
Research in Astronomy, Inc., under NASA contract NAS5--26555. Support for
MAST for non--HST data is provided by the NASA Office of Space Science
via grant NAG5--7584 and by other grants and contracts.  
This research has made use of the SIMBAD database, operated at CDS,
Strasbourg, France.  

{\it Facility:} \facility{GALEX}



\bibliographystyle{apj}
\bibliography{bibs}



\input{table1-long}  
\clearpage

\input{table2-long} 
\clearpage

\input{table3}  

\clearpage
\input{table4}  

\end{document}

%% file: table1-long.tex


\begin{deluxetable}{lll}
\tablewidth{0pt}
\tablecaption{PGR sources and 2MASS counterparts\label{table1}} 
\tablehead{
   \colhead{PGR Name} & \colhead{2MASS Designation} & \colhead{Notes}
 }

\startdata
PGR J00021+0251 & 2MASS J00020846+0251282 &      \\
PGR J00036+0013 & 2MASS J00033771+0013085 &      \\
PGR J00040+0251 & 2MASS J00040522+0251534 &      \\
PGR J00075+0542 & 2MASS J00073216+0542017 &      \\
PGR J00373+0628 & 2MASS J00371927+0628165 &      \\
PGR J00388+1228 & 2MASS J00385212+1228112 &      \\
PGR J00410+0157 & 2MASS J00410389+0157473 &      \\
PGR J00418+0345 & 2MASS J00415150+0345285 &  1   \\
PGR J00434+0704 & 2MASS J00432693+0704389 &      \\
PGR J00446+1055 & 2MASS J00443778+1055320 &      \\
PGR J00451+1802 & 2MASS J00450738+1802573 &      \\
PGR J00515+0021 & 2MASS J00513429+0021393 &      \\
PGR J00547+0224 & 2MASS J00544424+0224060 &      \\
PGR J00550+0905 & 2MASS J00550423+0905570 &      \\
PGR J00577+0631 & 2MASS J00574772+0631090 &  1   \\
PGR J00586+0431 & 2MASS J00583732+0431146 &  1   \\
PGR J00590+1255 & 2MASS J00590159+1255505 &      \\
PGR J01010+2230 & 2MASS J01010193+2230303 &      \\
PGR J01028+1324 & 2MASS J01025095+1324548 &  1   \\
PGR J01048+1639 & 2MASS J01045053+1639035 &      \\
PGR J01051+1624 & 2MASS J01050611+1624562 &      \\
PGR J01109+1845 & 2MASS J01105729+1845052 &      \\
PGR J01117+1746 & 2MASS J01114271+1746481 &      \\
PGR J01122+1813 & 2MASS J01121375+1813495 &      \\
PGR J01135+1845 & 2MASS J01133046+1845008 &      \\
PGR J01162+1714 & 2MASS J01161310+1714194 &      \\
PGR J01162+1840 & 2MASS J01161525+1840381 &      \\
PGR J01369+1653 & 2MASS J01365914+1653152 &      \\
PGR J02000+1409 & 2MASS J02000162+1409419 &  1   \\
PGR J02040+1500 & 2MASS J02040015+1500434 &      \\
PGR J02047+1514 & 2MASS J02044221+1514550 &      \\
PGR J02549+1259 & 2MASS J02545675+1259248 &      \\
PGR J02561+1257 & 2MASS J02560830+1257089 &      \\
PGR J02572+1313 & 2MASS J02571584+1313253 &      \\
PGR J03037+1202 & 2MASS J03034569+1202136 &      \\
PGR J03130+0313 & 2MASS J03130291+0314001 &      \\
PGR J03133+0542 & 2MASS J03131814+0542483 &      \\
PGR J03281+0035 & 2MASS J03280935+0035272 &  1   \\
PGR J08237+6750 & 2MASS J08234774+6750216 &      \\
PGR J08315+4047 & 2MASS J08313260+4047091 &      \\
PGR J08391+4710 & 2MASS J08390680+4710519 &      \\
PGR J08392+4624 & 2MASS J08391477+4624416 &      \\
PGR J08401+4421 & 2MASS J08400897+4421272 &      \\
PGR J08430+4447 & 2MASS J08430003+4447280 &      \\
PGR J08430+4609 & 2MASS J08430306+4609075 &      \\
PGR J08431+4606 & 2MASS J08430627+4606269 &  1,2 \\
PGR J08433+4606 & 2MASS J08432100+4606532 &      \\
PGR J08484+2714 & 2MASS J08482915+2714548 &      \\
PGR J08538+2708 & 2MASS J08534884+2708382 &      \\
PGR J09059+6027 & 2MASS J09055922+6027162 &  2   \\
PGR J09069+5722 & 2MASS J09065794+5722567 &      \\
PGR J09087+6024 & 2MASS J09084225+6024023 &      \\
PGR J09165+3213 & 2MASS J09163419+3213061 &      \\
PGR J09190+4314 & 2MASS J09190162+4314000 &      \\
PGR J09233+5716 & 2MASS J09232048+5716029 &      \\
PGR J09281+6503 & 2MASS J09281137+6503311 &      \\
PGR J09322+5613 & 2MASS J09321441+5613272 &      \\
PGR J09343+6347 & 2MASS J09342377+6347564 &      \\
PGR J09370+4228 & 2MASS J09370212+4228321 &      \\
PGR J09395+6353 & 2MASS J09393451+6353016 &      \\
PGR J10170+5439 & 2MASS J10170106+5439076 &      \\
PGR J10173+5052 & 2MASS J10172228+5052286 &  1   \\
PGR J10203+3912 & 2MASS J10201970+3912284 &  1   \\
PGR J10273+5758 & 2MASS J10272284+5758345 &  1   \\
PGR J10319+6212 & 2MASS J10315611+6212506 &      \\
PGR J10337+5505 & 2MASS J10334782+5505580 &      \\
PGR J10445+5445 & 2MASS J10443579+5445066 &      \\
PGR J10453+5824 & 2MASS J10452381+5824022 &      \\
PGR J21005+0221 & 2MASS J21003209+0221100 &  1   \\
PGR J21556+0603 & 2MASS J21553620+0603295 &  1   \\
PGR J22172+1900 & 2MASS J22171637+1900375 &      \\
PGR J22199+1828 & 2MASS J22195611+1828022 &      \\
PGR J22216+1723 & 2MASS J22214179+1723557 &      \\
PGR J22247+0827 & 2MASS J22244501+0827532 &      \\
PGR J22263+1657 & 2MASS J22262100+1657432 &  1   \\
PGR J22272+1721 & 2MASS J22271650+1721569 &      \\
PGR J22311+1648 & 2MASS J22311155+1648190 &  1   \\
PGR J22451+2134 & 2MASS J22450803+2134194 &      \\
PGR J22564+2256 & 2MASS J22562910+2256281 &  1,2 \\
PGR J22572+2215 & 2MASS J22571430+2215237 &      \\
PGR J23001+2512 & 2MASS J23000676+2512365 &      \\
PGR J23025+2602 & 2MASS J23023384+2602579 &      \\
PGR J23038+2731 & 2MASS J23034979+2731381 &      \\
PGR J23254+1236 & 2MASS J23252669+1236423 &      \\
PGR J23263+0259 & 2MASS J23262172+0259315 &  1   \\
PGR J23445+0204 & 2MASS J23443281+0204199 &  1   \\
PGR J23479+1048 & 2MASS J23475699+1048552 &  1   \\
PGR J23562+2613 & 2MASS J23561539+2613419 &      \\
\enddata

\tablenotetext{1}{Positional offset between USNO-A2 and 2MASS exceeds
      {1\farcs5} (80th percentile of offset size).}
\tablenotetext{2}{There is a second 2MASS object within {10\arcsec} of
      the USNO position.}

\end{deluxetable}

%% file: table2-long.tex

\begin{deluxetable}{lrlrlrcr@{}rrrrl}
\tablewidth{0pt}
\tabletypesize{\scriptsize}
\tablecaption{GALEX and V-band Photometry of PGR objects\label{table2}}
\tablehead{
    \colhead{Name} & \colhead{$F$} & \colhead{$\sigma(F)$\tablenotemark{a}} & \colhead{$N$} &
    \colhead{$\sigma(N)$} &
    \colhead{$V$} & \colhead{Ref\tablenotemark{b}} & \multicolumn{2}{c}{$(F\!-\!N)$\tablenotemark{a}} &
    \colhead{$(N\!-\!V)$} & \colhead{E$(\bv)$\tablenotemark{c}} & \colhead{$(N\!-\!V)_\circ$} &
    \colhead{Remarks} \\
    \colhead{ } & \colhead{(mag)} & \colhead{(mag)} & \colhead{(mag)} & \colhead{(mag)} & \colhead{(mag)} &
    \colhead{ } & \multicolumn{2}{c}{(mag)} & \colhead{(mag)} & \colhead{(mag)} & \colhead{(mag)} &
    \colhead{ }
  }

\startdata
PGR J00021+0251 &  21.97 & \nodata &  18.14 &  0.04 &  14.2  &  A  & $>$ &  $   3.83$ &  3.94 &  0.022 & $   3.83$ & \nodata \\
PGR J00036+0013 &  21.73 & \nodata &  19.03 &  0.08 &  14.8  &  A  & $>$ &  $   2.70$ &  4.23 &  0.030 & $   4.09$ & \nodata \\
PGR J00040+0251 &  21.75 & \nodata &  18.38 &  0.06 &  14.8  &  A  & $>$ &  $   3.37$ &  3.58 &  0.021 & $   3.48$ & \nodata \\
PGR J00075+0542 &  13.47 &    0.01 &  13.72 &  0.01 &  13.0  &  D  &     &  $  -0.25$ &  0.72 &  0.031 & $   0.57$ & hot   \\
PGR J00373+0628 &  22.03 & \nodata &  18.92 &  0.05 &  15.1  &  A  & $>$ &  $   3.11$ &  3.82 &  0.022 & $   3.71$ & \nodata \\
PGR J00388+1228 &  21.49 & \nodata &  19.69 &  0.11 &  15.7  &  A  & $>$ &  $   1.80$ &  3.99 &  0.085 & $   3.58$ & \nodata \\
PGR J00410+0157 &  21.75 & \nodata &  19.02 &  0.08 &  14.7  &  A  & $>$ &  $   2.73$ &  4.32 &  0.018 & $   4.23$ & \nodata \\
PGR J00418+0345 &  21.83 & \nodata &  17.76 &  0.04 &  15.0  &  A  & $>$ &  $   4.07$ &  2.76 &  0.020 & $   2.66$ & \nodata \\
PGR J00434+0704 &  21.58 & \nodata &  17.85 &  0.03 &  14.6  &  A  & $>$ &  $   3.73$ &  3.25 &  0.038 & $   3.07$ & \nodata \\
PGR J00446+1055 &  21.73 & \nodata &  19.33 &  0.09 &  14.8  &  A  & $>$ &  $   2.39$ &  4.53 &  0.061 & $   4.24$ & \nodata \\
PGR J00451+1802 &  21.83 & \nodata &  18.52 &  0.06 &  14.1  &  A  & $>$ &  $   3.31$ &  4.42 &  0.056 & $   4.15$ & \nodata \\
PGR J00515+0021 &  21.71 & \nodata &  19.79 &  0.13 &  15.9  &  B  & $>$ &  $   1.92$ &  3.89 &  0.022 & $   3.78$ & \nodata \\
PGR J00547+0224 &  21.64 & \nodata &  19.26 &  0.10 &  14.8  &  A  & $>$ &  $   2.38$ &  4.46 &  0.029 & $   4.32$ & \nodata \\
PGR J00550+0905 &  22.31 & \nodata &  19.67 &  0.11 &  14.6  &  A  & $>$ &  $   2.63$ &  5.07 &  0.055 & $   4.81$ & \nodata \\
PGR J00577+0631 &  21.62 & \nodata &  16.89 &  0.03 &  14.1  &  A  & $>$ &  $   4.73$ &  2.79 &  0.048 & $   2.56$ & \nodata \\
PGR J00586+0431 &  21.53 & \nodata &  17.13 &  0.02 &  14.8  &  A  & $>$ &  $   4.40$ &  2.33 &  0.024 & $   2.21$ & \nodata \\
PGR J00590+1255 &  21.50 & \nodata &  19.21 &  0.09 &  14.0  &  D  & $>$ &  $   2.30$ &  5.21 &  0.071 & $   4.87$ & \nodata \\
PGR J01010+2230 &  21.70 & \nodata &  18.23 &  0.03 &  13.0  &  D  & $>$ &  $   3.47$ &  5.23 &  0.035 & $   5.06$ & \nodata \\
PGR J01028+1324 &  23.21 & \nodata &  17.95 &  0.01 &  13.7  &  D  & $>$ &  $   5.27$ &  4.25 &  0.035 & $   4.08$ & MIS \\
PGR J01048+1639 &  22.41 & \nodata &  18.94 &  0.02 &  14.2  &  A  & $>$ &  $   3.46$ &  4.75 &  0.049 & $   4.51$ & GI \\
PGR J01051+1624 &  22.58 & \nodata &  17.70 &  0.01 &  13.5  &  D  & $>$ &  $   4.87$ &  4.20 &  0.050 & $   3.96$ & GI  \\
PGR J01109+1845 &  21.91 & \nodata &  18.79 &  0.05 &  14.5  &  A  & $>$ &  $   3.12$ &  4.29 &  0.048 & $   4.06$ & \nodata \\
PGR J01117+1746 &  21.72 & \nodata &  18.78 &  0.07 &  14.3  &  A  & $>$ &  $   2.94$ &  4.48 &  0.077 & $   4.11$ & \nodata \\
PGR J01122+1813 &  21.44 & \nodata &  17.43 &  0.03 &  13.7  &  D  & $>$ &  $   4.01$ &  3.73 &  0.060 & $   3.44$ & \nodata \\
PGR J01135+1845 &  22.04 & \nodata &  20.36 &  0.17 &  15.0  &  A  & $>$ &  $   1.67$ &  5.36 &  0.047 & $   5.13$ & \nodata \\
PGR J01162+1714 &  21.53 & \nodata &  16.70 &  0.02 &  12.2  &  D  & $>$ &  $   4.84$ &  4.50 &  0.071 & $   4.16$ & \nodata \\
PGR J01162+1840 &  21.85 & \nodata &  18.28 &  0.05 &  15.1  &  A  & $>$ &  $   3.57$ &  3.18 &  0.050 & $   2.94$ & \nodata \\
PGR J01369+1653 &  21.51 & \nodata &  19.78 &  0.12 &  15.2  &  A  & $>$ &  $   1.74$ &  4.58 &  0.110 & $   4.05$ & \nodata \\
PGR J02000+1409 &  20.62 &    0.07 &  15.72 &  0.01 &  13.0  &  D  &     &  $   4.90$ &  2.72 &  0.053 & $   2.47$ & MIS \\
PGR J02040+1500 &  21.90 &    0.47 &  18.74 &  0.06 &  14.3  &  A  &     &  $   3.16$ &  4.44 &  0.076 & $   4.07$ & comp? \\
PGR J02047+1514 &  22.41 &    0.14 &  15.71 &  0.01 &  11.6  &  D  &     &  $   6.70$ &  4.11 &  0.051 & $   3.87$ & GI \\
PGR J02549+1259 &  21.43 & \nodata &  19.03 &  0.08 &  14.7  &  A  & $>$ &  $   2.40$ &  4.33 &  0.183 & $   3.45$ & \nodata \\
PGR J02561+1257 &  21.36 & \nodata &  19.49 &  0.08 &  15.5  &  A  & $>$ &  $   1.87$ &  3.99 &  0.164 & $   3.20$ & \nodata \\
PGR J02572+1313 &  21.52 & \nodata &  18.96 &  0.06 &  14.6  &  A  & $>$ &  $   2.56$ &  4.36 &  0.148 & $   3.65$ & \nodata \\
PGR J03037+1202 &  21.16 & \nodata &  18.71 &  0.07 &  14.4  &  A  & $>$ &  $   2.45$ &  4.31 &  0.264 & $   3.04$ & \nodata \\
PGR J03130+0313 &  21.07 & \nodata &  17.91 &  0.05 &  13.5  &  D  & $>$ &  $   3.15$ &  4.41 &  0.095 & $   3.95$ & \nodata \\
PGR J03133+0542 &  21.37 & \nodata &  19.30 &  0.07 &  14.1  &  A  & $>$ &  $   2.07$ &  5.20 &  0.233 & $   4.08$ & \nodata \\
PGR J03281+0035 &  21.13 & \nodata &  17.39 &  0.04 &  14.3  &  A  & $>$ &  $   3.75$ &  3.09 &  0.123 & $   2.50$ & \nodata \\
PGR J08237+6750 &  21.46 &    0.38 &  16.48 &  0.02 &  13.9  &  A  &     &  $   4.97$ &  2.58 &  0.042 & $   2.38$ & \nodata \\
PGR J08315+4047 &  20.84 &    0.31 &  15.97 &  0.02 &  12.8  &  A  &     &  $   4.87$ &  3.17 &  0.039 & $   2.98$ & \nodata \\
PGR J08391+4710 &  22.81 & \nodata &  17.75 &  0.01 &  13.8  &  A  & $>$ &  $   5.07$ &  3.95 &  0.027 & $   3.82$ & MIS \\
PGR J08392+4624 &  21.51 & \nodata &  16.62 &  0.02 &  13.0  &  A  & $>$ &  $   4.89$ &  3.62 &  0.026 & $   3.49$ & \nodata \\
PGR J08401+4421 &  20.65 &    0.23 &  16.88 &  0.03 &  13.2  &  A  &     &  $   3.77$ &  3.68 &  0.024 & $   3.57$ & comp? \\
PGR J08430+4447 &  21.33 & \nodata &  16.24 &  0.01 &  12.9  &  A  & $>$ &  $   5.09$ &  3.34 &  0.026 & $   3.21$ & \nodata \\
PGR J08430+4609 &  20.98 &    0.28 &  16.60 &  0.02 &  14.1  &  A  &     &  $   4.39$ &  2.50 &  0.028 & $   2.37$ & \nodata \\
PGR J08431+4606 &  22.09 & \nodata &  16.06 &  0.02 &  12.4  &  A  & $>$ &  $   6.04$ &  3.65 &  0.029 & $   3.51$ & \nodata \\
PGR J08433+4606 &  21.00 &    0.26 &  16.44 &  0.02 &  13.5  &  A  &     &  $   4.57$ &  2.94 &  0.029 & $   2.80$ & \nodata \\
PGR J08484+2714 &  21.69 & \nodata &  18.78 &  0.06 &  15.3  &  B  & $>$ &  $   2.92$ &  3.48 &  0.047 & $   3.25$ & \nodata \\
PGR J08538+2708 &  21.57 & \nodata &  20.12 &  0.15 &  15.3  &  B  & $>$ &  $   1.44$ &  4.82 &  0.033 & $   4.66$ & \nodata \\
PGR J09059+6027 &  22.87 & \nodata &  19.83 &  0.11 &  16.3  &  B  & $>$ &  $   3.04$ &  3.52 &  0.035 & $   3.35$ & \nodata \\
PGR J09069+5722 &  22.95 & \nodata &  18.28 &  0.01 &  13.6  &  B  & $>$ &  $   4.67$ &  4.68 &  0.026 & $   4.55$ & MIS \\
PGR J09087+6024 &  22.90 & \nodata &  19.91 &  0.03 &  15.4  &  B  & $>$ &  $   2.99$ &  4.51 &  0.039 & $   4.32$ & NGA \\
PGR J09165+3213 &  20.88 &    0.22 &  15.40 &  0.01 &  12.4  &  A  &     &  $   5.48$ &  3.00 &  0.019 & $   2.91$ & \nodata \\
PGR J09190+4314 &  21.68 & \nodata &  17.13 &  0.02 &  14.1  &  A  & $>$ &  $   4.55$ &  3.03 &  0.014 & $   2.96$ & \nodata \\
PGR J09233+5716 &  22.57 & \nodata &  17.99 &  0.01 &  12.7  &  A  & $>$ &  $   4.57$ &  5.29 &  0.036 & $   5.12$ & MIS \\
PGR J09281+6503 &  21.94 &    0.47 &  16.23 &  0.01 &  12.5  &  C  &     &  $   5.71$ &  3.73 &  0.062 & $   3.43$ & \nodata \\
PGR J09322+5613 &  21.82 & \nodata &  17.76 &  0.04 &  13.1  &  A  & $>$ &  $   4.06$ &  4.66 &  0.023 & $   4.55$ & \nodata \\
PGR J09343+6347 &  21.80 & \nodata &  18.61 &  0.06 &  15.4  &  B  & $>$ &  $   3.18$ &  3.21 &  0.031 & $   3.06$ & \nodata \\
PGR J09370+4228 &  22.27 & \nodata &  17.38 &  0.03 &  13.5  &  A  & $>$ &  $   4.90$ &  3.88 &  0.013 & $   3.82$ & \nodata \\
PGR J09395+6353 &  21.72 & \nodata &  20.08 &  0.13 &  15.0  &  B  & $>$ &  $   1.64$ &  5.08 &  0.037 & $   4.90$ & \nodata \\
PGR J10170+5439 &  21.70 & \nodata &  18.05 &  0.05 &  14.1  &  B  & $>$ &  $   3.65$ &  3.95 &  0.008 & $   3.91$ & \nodata \\
PGR J10173+5052 &  21.14 & \nodata &  17.61 &  0.04 &  14.0  &  A  & $>$ &  $   3.53$ &  3.61 &  0.010 & $   3.56$ & \nodata \\
PGR J10203+3912 &  22.14 & \nodata &  17.45 &  0.02 &  13.2  &  A  & $>$ &  $   4.69$ &  4.25 &  0.012 & $   4.19$ & \nodata \\
PGR J10273+5758 &  23.54 & \nodata &  17.37 &  0.01 &  13.4  &  C  & $>$ &  $   6.17$ &  3.97 &  0.014 & $   3.90$ & LOCK \\
PGR J10319+6212 &  23.40 &    0.32 &  17.35 &  0.01 &  14.0  &  B  &     &  $   6.05$ &  3.35 &  0.014 & $   3.28$ & GI \\
PGR J10337+5505 &  21.58 &    0.29 &  15.44 &  0.01 &  12.6  &  A  &     &  $   6.14$ &  2.84 &  0.008 & $   2.80$ & \nodata \\
PGR J10445+5445 &  22.23 & \nodata &  18.54 &  0.06 &  13.1  &  A  & $>$ &  $   3.69$ &  5.44 &  0.011 & $   5.39$ & \nodata \\
PGR J10453+5824 &  24.35 &    0.17 &  17.79 &  0.01 &  13.8  &  A  &     &  $   6.56$ &  3.99 &  0.010 & $   3.94$ & LOCK \\
PGR J21005+0221 &  21.70 & \nodata &  18.99 &  0.06 &  15.7  &  A  & $>$ &  $   2.71$ &  3.29 &  0.098 & $   2.82$ & \nodata \\
PGR J21556+0603 &  21.34 & \nodata &  17.82 &  0.04 &  14.8  &  A  & $>$ &  $   3.52$ &  3.02 &  0.051 & $   2.78$ & \nodata \\
PGR J22172+1900 &  21.85 & \nodata &  18.94 &  0.08 &  14.8  &  A  & $>$ &  $   2.90$ &  4.14 &  0.048 & $   3.91$ & \nodata \\
PGR J22199+1828 &  21.51 & \nodata &  18.33 &  0.06 &  14.1  &  A  & $>$ &  $   3.17$ &  4.23 &  0.046 & $   4.01$ & \nodata \\
PGR J22216+1723 &  21.32 & \nodata &  18.87 &  0.08 &  14.9  &  A  & $>$ &  $   2.44$ &  3.97 &  0.049 & $   3.73$ & \nodata \\
PGR J22247+0827 &  21.72 & \nodata &  19.31 &  0.09 &  14.4  &  A  & $>$ &  $   2.41$ &  4.91 &  0.096 & $   4.45$ & \nodata \\
PGR J22263+1657 &  21.35 & \nodata &  17.89 &  0.03 &  15.3  &  A  & $>$ &  $   3.46$ &  2.59 &  0.058 & $   2.31$ & \nodata \\
PGR J22272+1721 &  21.16 & \nodata &  17.52 &  0.04 &  15.0  &  A  & $>$ &  $   3.64$ &  2.52 &  0.066 & $   2.20$ & \nodata \\
PGR J22311+1648 &  21.27 & \nodata &  17.47 &  0.03 &  14.8  &  A  & $>$ &  $   3.81$ &  2.66 &  0.071 & $   2.32$ & \nodata \\
PGR J22451+2134 &  14.47 &    0.01 &  14.58 &  0.01 &  14.4  &  A  &     &  $  -0.11$ &  0.18 &  0.057 & $  -0.09$ & hot   \\
PGR J22564+2256 &  21.79 & \nodata &  18.32 &  0.06 &  15.3  &  A  & $>$ &  $   3.47$ &  3.02 &  0.047 & $   2.79$ & \nodata \\
PGR J22572+2215 &  21.45 & \nodata &  19.52 &  0.08 &  15.2  &  A  & $>$ &  $   1.93$ &  4.32 &  0.043 & $   4.11$ & \nodata \\
PGR J23001+2512 &  22.17 & \nodata &  20.16 &  0.14 &  14.9  &  A  & $>$ &  $   2.01$ &  5.26 &  0.107 & $   4.75$ & \nodata \\
PGR J23025+2602 &  15.06 &    0.01 &  15.33 &  0.01 &  15.1  &  A  &     &  $  -0.27$ &  0.23 &  0.075 & $  -0.13$ & hot   \\
PGR J23038+2731 &  21.40 & \nodata &  18.47 &  0.06 &  14.7  &  A  & $>$ &  $   2.93$ &  3.77 &  0.059 & $   3.49$ & \nodata \\
PGR J23254+1236 &  22.01 & \nodata &  20.26 &  0.12 &  15.3  &  A  & $>$ &  $   1.74$ &  4.96 &  0.065 & $   4.65$ & \nodata \\
PGR J23263+0259 &  21.52 & \nodata &  19.95 &  0.14 &  15.5  &  A  & $>$ &  $   1.57$ &  4.45 &  0.040 & $   4.26$ & \nodata \\
PGR J23445+0204 &  21.19 & \nodata &  18.30 &  0.06 &  15.7  &  A  & $>$ &  $   2.89$ &  2.60 &  0.035 & $   2.43$ & \nodata \\
PGR J23479+1048 &  21.20 & \nodata &  16.71 &  0.03 &  13.8  &  D  & $>$ &  $   4.50$ &  2.91 &  0.051 & $   2.66$ & \nodata \\
PGR J23562+2613 &  21.57 & \nodata &  18.08 &  0.04 &  15.1  &  A  & $>$ &  $   3.49$ &  2.98 &  0.036 & $   2.81$ & \nodata \\
\enddata

\tablenotetext{a}{If no error is reported, then the $F$ magnitude is an upper limit --- in this case
                  the $(F\!-\!N)$ color is also a limit and is preceded by a ``$>$'' symbol.}
\tablenotetext{b}{Reference for the adopted $V$ magnitude (see \S\ref{sec:Visual}).}
\tablenotetext{c}{From \citet{Schlegel}, as returned by the MAST query.}
\tablenotetext{d}{Dereddened $(N\!-\!V)$ color, using E$(N\!-\!V)=4.8\times\mathrm{E}(\bv)$.}
\end{deluxetable}

%% file: table3.tex

\begin{deluxetable}{llrrrrrr}
\tabletypesize{\scriptsize}
\tablecaption{Galex and other measurements for known hot subdwarfs.\label{table3}}
\tablehead{
    \colhead{Name} & \colhead{GALEX Name} & \colhead{$F$} & \colhead{$N$} & \colhead{{$V$}\tablenotemark{1}} &
    \colhead{E$(\bv)$} & \colhead{$(F\!-\!N)_0$} & \colhead{$(N\!-\!V)_0$}  \\ 
    \colhead{ } & \colhead{ } & \colhead{(mag)} & \colhead{(mag)} & \colhead{(mag)} & \colhead{(mag)} &
       \colhead{(mag)} & \colhead{(mag)} 
  }

\startdata
 PG 0105+276\tablenotemark{2} & GALEX J010816.5+275253 & 13.97 & 13.72 &  14.45 & 0.058 & $+0.25$ &  $-1.00$ \\
 PG 0212+148                  & GALEX J021511.2+150005 & 13.74 & 13.98 &  14.45 & 0.088 & $-0.23$ &  $-0.90$ \\
 PG 0220+132                  & GALEX J022338.4+132734 & 14.08 & 14.40 &  14.78 & 0.122 & $-0.31$ &  $-0.96$ \\
 TON 349                      & GALEX J084718.9+230030 & 13.87 & 13.92 &  15.34 & 0.031 & $-0.04$ &  $-1.57$ \\
 PG 2335+107                  & GALEX J233743.8+105627 & 14.63 & 14.77 &  15.55 & 0.061 & $-0.13$ &  $-1.07$ \\
 PG 2356+167                  & GALEX J235925.3+165641 & 13.63 & 13.93 &  14.21 & 0.044 & $-0.30$ &  $-0.49$ \\
\enddata

\tablenotetext{1}{$V$ magnitudes from
\protect{\citet{Allard}} (PG 0105+276, PG 0212+148, PG 0220+132)
and \protect{\citet{PG}} (TON 349); Str\"omgren
$y$ magnitudes from \protect{\citet{PG}} (PG 2335+107) 
and \protect{\citet{Wesemael}} (PG 2356+167).}

\tablenotetext{2}{Visual double, see \S\ref{sec:Visual}.}

\end{deluxetable}

%% file: table4.tex

\begin{deluxetable}{lccccccl}
\tablewidth{0pt}
\tabletypesize{\scriptsize}
\tablecaption{GALEX and 2MASS Photometry of Selected PGR objects\label{table4}}
\tablehead{
    \colhead{Name} & \colhead{$(F\!-\!N)$} & \colhead{$(N\!-\!V)_\circ$} &
    \colhead{$V$} & \colhead{$(V\!-\!K_s)$} & \colhead{$(J\!-\!H)$} & \colhead{$(J\!-\!K_s)$} & 
    \colhead{Remarks} \\
    \colhead{ } & \colhead{(mag)} & \colhead{(mag)} & \colhead{(mag)} & \colhead{(mag)} & \colhead{(mag)} &
    \colhead{(mag)} & \colhead{ }
  }
\startdata
PGR J00075+0542 & $-0.25\pm0.01$ & $+0.57\pm0.05$ & $13.0\pm0.05$ & $+0.7\pm0.05$ & $0.15\pm0.03$ & $0.18\pm0.03$ & hot   \\
PGR J02040+1500 & $+3.16\pm0.47$ & $+4.07\pm0.26$ & $14.3\pm0.25$ & $+1.9\pm0.25$ & $0.36\pm0.04$ & $0.37\pm0.04$ & comp? \\
PGR J08401+4421 & $+3.77\pm0.23$ & $+3.57\pm0.25$ & $13.2\pm0.25$ & $+1.6\pm0.25$ & $0.26\pm0.04$ & $0.34\pm0.04$ & comp? \\
PGR J22451+2134 & $-0.11\pm0.01$ & $-0.09\pm0.25$ & $14.4\pm0.25$ & $+2.0\pm0.25$ & $0.40\pm0.03$ & $0.45\pm0.03$ & hot   \\
PGR J23025+2602 & $-0.27\pm0.01$ & $-0.13\pm0.25$ & $15.1\pm0.25$ & $+1.6\pm0.25$ & $0.21\pm0.04$ & $0.27\pm0.04$ & hot   \\
\enddata
\end{deluxetable}